\input harvmac
\input epsf
\input color
\input amssym
\font\small =cmr10 scaled 850
\def\vev#1{\langle#1\rangle}
\def\frac#1#2{{{#1}\over{#2}}}

\def\centertable#1{ \hbox to \hsize {\hfill\vbox{
                    \offinterlineskip \tabskip=0pt \halign{#1} }\hfill} }

\def\wt{\widetilde}

\def\eq{\eqn\nolabel}

\def\L{{\cal L}}

\def\J{{\cal J}}
\def\D{{\cal D}}

\def\t{\theta}

\def\date{\number\month/\number\day/\number\yearltd}

\def\centertable#1{ \hbox to \hsize {\hfill\vbox{
                    \offinterlineskip \tabskip=0pt \halign{#1} }\hfill} }

\lref\dgp{
  S.~Dimopoulos, G.~F.~Giudice and A.~Pomarol,
  ``Dark matter in theories of gauge-mediated supersymmetry breaking,''
  Phys.\ Lett.\  B {\bf 389}, 37 (1996)
  [arXiv:hep-ph/9607225].
}                

\lref\martin{
  S.~P.~Martin,
  ``Generalized messengers of supersymmetry breaking and the sparticle mass
  spectrum,''
  Phys.\ Rev.\  D {\bf 55}, 3177 (1997)
  [arXiv:hep-ph/9608224].
}

\lref\hgm{
  E.~Gorbatov and M.~Sudano,
  ``Sparticle Masses in Higgsed Gauge Mediation,''
  JHEP {\bf 0810}, 066 (2008)
  [arXiv:0802.0555 [hep-ph]].
}        

\lref\comments{
  K.~A.~Intriligator and M.~Sudano,
  ``Comments on General Gauge Mediation,''
  JHEP {\bf 0811}, 008 (2008)
  [arXiv:0807.3942 [hep-ph]].
}

\lref\ggmgm{
  K.~Intriligator and M.~Sudano,
  ``General Gauge Mediation with Gauge Messengers,''
  JHEP {\bf 1006}, 047 (2010)
  [arXiv:1001.5443 [hep-ph]].
}

\lref\Buican{
  M.~Buican and Z.~Komargodski,
  ``Soft Terms from Broken Symmetries,''
  JHEP {\bf 1002}, 005 (2010)
  [arXiv:0909.4824 [hep-ph]].
}

\lref\Kap{
  D.~E.~Kaplan, G.~D.~Kribs and M.~Schmaltz,
  ``Supersymmetry breaking through transparent extra dimensions,''
  Phys.\ Rev.\  D {\bf 62}, 035010 (2000)
  [arXiv:hep-ph/9911293].
}

\lref\Chack{
  Z.~Chacko, M.~A.~Luty, A.~E.~Nelson and E.~Ponton,
  ``Gaugino mediated supersymmetry breaking,''
  JHEP {\bf 0001}, 003 (2000)
  [arXiv:hep-ph/9911323].
}

\lref\Csaki{
  C.~Csaki, J.~Erlich, C.~Grojean and G.~D.~Kribs,
  ``4D constructions of supersymmetric extra dimensions and gaugino
  mediation,''
  Phys.\ Rev.\  D {\bf 65}, 015003 (2002)
  [arXiv:hep-ph/0106044].
}

\lref\Cheng{
  H.~C.~Cheng, D.~E.~Kaplan, M.~Schmaltz and W.~Skiba,
  ``Deconstructing gaugino mediation,''
  Phys.\ Lett.\  B {\bf 515}, 395 (2001)
  [arXiv:hep-ph/0106098].
}

\lref\Green{
  D.~Green, A.~Katz and Z.~Komargodski,
  ``Direct Gaugino Mediation,''
  arXiv:1008.2215 [hep-th].
}

\lref\Meade{
  P.~Meade, N.~Seiberg and D.~Shih,
  ``General Gauge Mediation,''
  Prog.\ Theor.\ Phys.\ Suppl.\  {\bf 177}, 143 (2009)
  [arXiv:0801.3278 [hep-ph]].
}

\lref\DineZA{
  M.~Dine, W.~Fischler and M.~Srednicki,
  ``Supersymmetric Technicolor,''
  Nucl.\ Phys.\  B {\bf 189}, 575 (1981).
}
\lref\DimAU{
  S.~Dimopoulos and S.~Raby,
  ``Supercolor,''
  Nucl.\ Phys.\  B {\bf 192}, 353 (1981).
}
\lref\WittenKV{
  E.~Witten,
  ``Mass Hierarchies In Supersymmetric Theories,''
  Phys.\ Lett.\  B {\bf 105}, 267 (1981).
}
\lref\NappiHM{
  C.~R.~Nappi and B.~A.~Ovrut,
  ``Supersymmetric Extension Of The SU(3) X SU(2) X U(1) Model,''
  Phys.\ Lett.\  B {\bf 113}, 175 (1982).
}
\lref\Alvarez{
  L.~Alvarez-Gaume, M.~Claudson and M.~B.~Wise,
  ``Low-Energy Supersymmetry,''
  Nucl.\ Phys.\  B {\bf 207}, 96 (1982).
}
\lref\DimGM{
  S.~Dimopoulos and S.~Raby,
  ``Geometric Hierarchy,''
  Nucl.\ Phys.\  B {\bf 219}, 479 (1983).
}
\lref\ADS{
  I.~Affleck, M.~Dine and N.~Seiberg,
  ``Dynamical Supersymmetry Breaking In Four-Dimensions And Its
  Phenomenological Implications,''
  Nucl.\ Phys.\  B {\bf 256}, 557 (1985).
}
\lref\DineYW{
  M.~Dine and A.~E.~Nelson,
  ``Dynamical supersymmetry breaking at low-energies,''
  Phys.\ Rev.\  D {\bf 48}, 1277 (1993)
  [arXiv:hep-ph/9303230].
}
\lref\DineVC{
  M.~Dine, A.~E.~Nelson and Y.~Shirman,
  ``Low-Energy Dynamical Supersymmetry Breaking Simplified,''
  Phys.\ Rev.\  D {\bf 51}, 1362 (1995)
  [arXiv:hep-ph/9408384].
}
\lref\DineAG{
  M.~Dine, A.~E.~Nelson, Y.~Nir and Y.~Shirman,
  ``New tools for low-energy dynamical supersymmetry breaking,''
  Phys.\ Rev.\  D {\bf 53}, 2658 (1996)
  [arXiv:hep-ph/9507378].
}
\lref\DineXK{
  M.~Dine, Y.~Nir and Y.~Shirman,
  ``Variations on minimal gauge mediated supersymmetry breaking,''
  Phys.\ Rev.\  D {\bf 55}, 1501 (1997)
  [arXiv:hep-ph/9607397].
}
\lref\GiudiceBP{
  G.~F.~Giudice and R.~Rattazzi,
  ``Theories with gauge-mediated supersymmetry breaking,''
  Phys.\ Rept.\  {\bf 322}, 419 (1999)
  [arXiv:hep-ph/9801271].
}

\lref\PolchinskiAN{
  J.~Polchinski and L.~Susskind,
  ``Breaking Of Supersymmetry At Intermediate-Energy,''
  Phys.\ Rev.\  D {\bf 26}, 3661 (1982).
}

\lref\IzawaGS{
  K.~I.~Izawa, Y.~Nomura, K.~Tobe and T.~Yanagida,
  ``Direct-transmission models of dynamical supersymmetry breaking,''
  Phys.\ Rev.\  D {\bf 56}, 2886 (1997)
  [arXiv:hep-ph/9705228].
}

\lref\ShiraiRR{
  S.~Shirai, M.~Yamazaki and K.~Yonekura,
  ``Aspects of Non-minimal Gauge Mediation,''
  JHEP {\bf 1006}, 056 (2010)
  [arXiv:1003.3155 [hep-ph]].
}

\lref\KomargodskiJF{
  Z.~Komargodski and D.~Shih,
  ``Notes on SUSY and R-Symmetry Breaking in Wess-Zumino Models,''
  JHEP {\bf 0904}, 093 (2009)
  [arXiv:0902.0030 [hep-th]].
}

\lref\giveon{
  R.~Auzzi and A.~Giveon,
  ``The sparticle spectrum in Minimal gaugino-Gauge Mediation,''
  arXiv:1009.1714 [hep-ph].
}

\lref\McGa{
  M.~McGarrie and R.~Russo,
  ``General Gauge Mediation in 5D,''
  Phys.\ Rev.\  D {\bf 82}, 035001 (2010)
  [arXiv:1004.3305 [hep-ph]].
}

\lref\McGb{
  M.~McGarrie,
  ``General Gauge Mediation and Deconstruction,''
  arXiv:1009.0012 [hep-ph].
}

\lref\deca{
  N.~Arkani-Hamed, A.~G.~Cohen and H.~Georgi,
  ``(De)constructing dimensions,''
  Phys.\ Rev.\ Lett.\  {\bf 86}, 4757 (2001)
  [arXiv:hep-th/0104005].
}

\lref\decb{
  C.~T.~Hill, S.~Pokorski and J.~Wang,
  ``Gauge invariant effective Lagrangian for Kaluza-Klein modes,''
  Phys.\ Rev.\  D {\bf 64}, 105005 (2001)
  [arXiv:hep-th/0104035].
}

\lref\mura{
  A.~de Gouvea, T.~Moroi and H.~Murayama,
  ``Cosmology of supersymmetric models with low-energy gauge mediation,''
  Phys.\ Rev.\  D {\bf 56}, 1281 (1997)
  [arXiv:hep-ph/9701244].
}

\def\smallg{\refs{\PolchinskiAN\IzawaGS\ShiraiRR-\KomargodskiJF}}

\def\gmed{\refs{\DineZA\DimAU\WittenKV\NappiHM\Alvarez\DimGM\ADS\DineYW\DineVC\DineAG\DineXK-\GiudiceBP}}

\Title{\vbox{\baselineskip12pt \hbox{IPMU 10-0155}}}
{\vbox{\centerline{General Gaugino Mediation}}}
\medskip

\centerline{\it Matthew Sudano}
\bigskip

\centerline{Institute for the Physics and Mathematics of the
Universe} \centerline{University of Tokyo} \centerline{Kashiwa,
Chiba 277-8583, Japan}

\smallskip

\vglue .3cm

\bigskip
\vskip 1cm

\noindent 

The spectrum of a class of gaugino mediation models with arbitrary hidden sector is considered.  These models are defined by a diagonal breaking of the mediating gauge group, which places them outside the realm of General Gauge Mediation.  While gauginos get masses as in ordinary gauge mediation, the scalar masses are screened.

\bigskip

\Date{September 2010}

\newsec{Introduction}

In this short note, I evaluate the soft spectrum resulting from a broad class of gauge-mediated \gmed\ supersymmetry (SUSY) breaking models that fall under the umbrella of gaugino mediation \refs{\Kap,\Chack}.  The class of models is not described by the standard formulas of general gauge mediation \Meade\ because a non-standard mediating gauge structure is assumed.  As we will discuss, the structure of the theories leads to a screening of scalar masses.  This is particularly appealing in the context of gauge mediation where complete models tend to yield a hierarchy between scalars and gauginos \smallg.   

The calculation is performed for four-dimensional models of the form described in \refs{\Cheng,\Csaki}.  Simple and explicit dynamical realizations of such constructions were recently found \Green.  We will consider a general SUSY-breaking hidden sector, expressing the results in terms of current correlators.  The result is shown to be expressible in terms of the elementary results for mediation via a massless \Meade\ or massive \comments\ vector multiplet.  

{\it Note added: The results of an explicit two-loop calculation of the spectrum in the minimal scenario were recently reported \giveon.}

\newsec{The Basic Setup}

Let's begin by considering a toy theory with gauge group $G=U(1)_1\times U(1)_2$ and with the following matter content.
\vskip.2in
\centertable{
\vrule height2.75ex depth1.25ex width 0.6pt #\tabskip=1em &
\hfil $#$\hfil&#\vrule&\hfil$#$\hfil&#\vrule&\hfil$#$\hfil&# \vrule width 0.6pt 
\tabskip=0pt
\cr
\noalign{\hrule height 0.6pt}
&&&{U(1)_1}&&{U(1)_2}&\cr
\noalign{\hrule}
&Q&&1&&0&\cr
&\wt Q&&-1&&0&\cr
&L&&1&&1&\cr
&\wt L&&-1&&-1&\cr
&\Phi&&0&&1&\cr
&\wt\Phi&&0&&-1&\cr
\noalign{\hrule height 0.6pt}
}
\noindent Let the superpotential be given by
\eqn\super{W=S\wt \Phi\Phi,}
and let the fields have non-zero vevs as follows.
\eqn\vevs{\vev S=M+\t^2F,\qquad\vev{\wt L}=\vev{L}=v.}
It is easy to generate these vevs from a complete theory, but those details are irrelevant for the present analysis.  

This model is similar to minimal gauge mediation, where the $Q$ fields would belong to the visible sector and the $\Phi$ would be messengers.  In minimal gauge mediation, these two sets of fields would interact through a common gauge group leading to two-loop visible-sector scalar masses.  Before the Higgsing, however, the ``selectron'' fields only interact with the $U(1)_1$ gauge fields, which interact with the ``link'' fields $L$, which interact with the $U(1)_2$ gauge fields, which finally interact with the messengers yielding scalar masses at four loops.  

After Higgsing the situation is quite different.  The vevs of the link fields break one linear combination of the $U(1)$'s while preserving another.  Since the selectron and messenger fields are both charged under these linear combinations, we essentially recover ordinary gauge mediation.  The difference is that one now has contributions via a combination of massless and massive gauge fields.  Note that one linear combination of gauginos is massive at tree level, while the other receives the ordinary one-loop mass \refs{\Meade,\martin}, so no new computation is needed here.   

\newsec{The Mass Calculation}

To evaluate the scalar masses, it is convenient to employ some of the methods of general gauge mediation \Meade.  We begin by writing part of the Lagrangian in superspace,
\eqn\jv{\L\supset2\int d^4\t(g_1\J_1V_1+g_2\J_2V_2).}
We continue to consider the simple gauge group, $G=U(1)_1\times U(1)_2$ where $V_{1,2}$ are the corresponding vector superfields and $g_{1,2}$ are the gauge coupling constants.  It's only an exercise in notation to generalize to realistic gauge groups.  The hidden sector, however, is now unconstrained (up to obvious assumptions needed to preserve the breaking structure).  In particular, we have
\eqn\jj{\J_1=Q^\dagger Q-\wt Q^\dagger\wt Q+L^\dagger L-\wt L^\dagger\wt L,\qquad\J_2=\J_h+L^\dagger L-\wt L^\dagger\wt L.} 
Under the shift to the stable vacuum, $L,\wt L\rightarrow L,\wt L+v$ ($v$ chosen positive without loss of generality), the Lagrangian picks up a term,
\eqn\eat{\L\supset4v\int d^4\t(g_1V_1+g_2V_2)\Sigma,}
\def\size{2}
\vskip-.2in
\centertable{
\vrule height.3ex depth0.25ex width 0pt \tabskip=1em  \hfil#\hfil\cr
\epsfxsize=\size truein\epsfbox{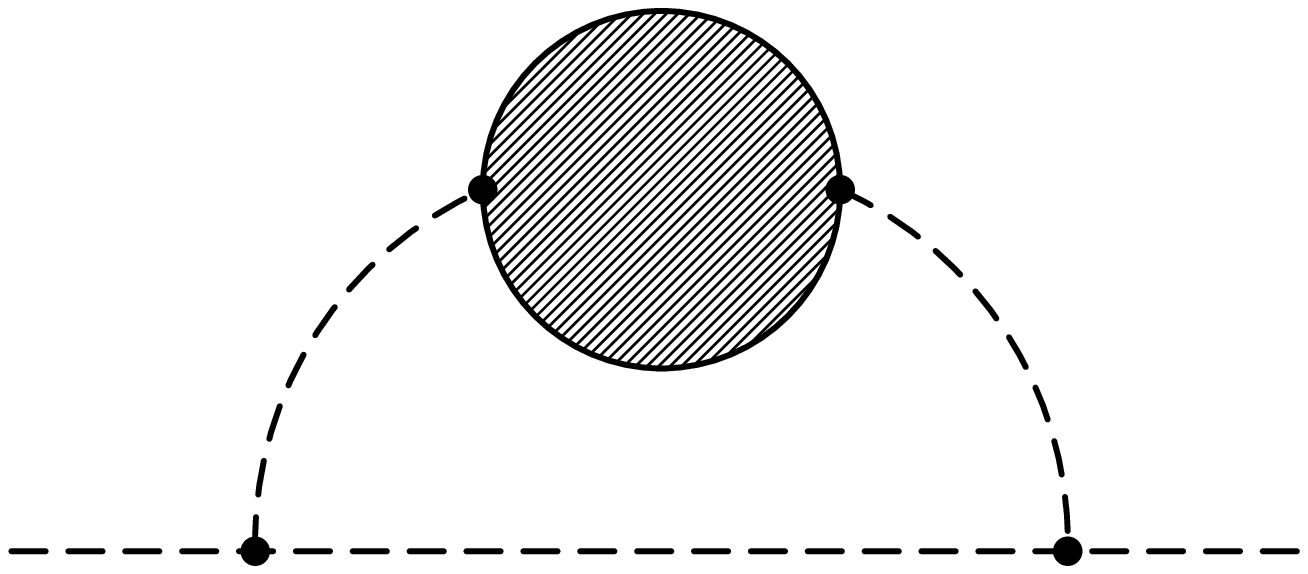}\cr
\cr
}
{\small\noindent{\bf Figure 1.} A term contributing to the mass of a visible sector scalar.  The propagator connecting this scalar to the hidden sector is $\vev{D_1D_2}$. }
\vskip.2in
\noindent
where $\Sigma\sim\hbox{Re}(L-\wt L)$ is the scalar component of the massive vector multiplet.  

The coupling \eat\ is all we need.  Consider the $D$-mediated contribution to the scalar component of $Q$ shown in Figure 1.  Since $Q$ only couples to $D_1$ and $\J_h$ only couples to $D_2$, the only relevant propagator is $\vev{D_1(p)D_2(-p)}$, which is determined by inverting a simple $3\times3$ matrix\foot{It is also straightforward to evaluate the full effective potential as in \comments.  In that case one needs a diagonal entry, $V(|Q|^2)=g_2^2\Tr\int\frac{d^4p}{(2\pi)^4}\vev{D_2(p)D_2(-p)}\left(\wt C_0(p^2)-4\wt C_{1/2}(p^2)+3\wt C_1(p^2)\right)$, evaluated with the visible-sector field given a background value.  The quartic and higher terms are generally irrelevant for collider physics, but they may be important for cosmology \mura.}:
\eq{\L\supset\frac12\D^T\Delta^{-1}\D,\qquad\D^T=(D_1,D_2,\Sigma)}
\eqn\propmat{\Delta^{-1}=\pmatrix{
1&0&2g_1v\cr
0&1&2g_2v\cr 
2g_1v&2g_2v&-p^2}\qquad\Rightarrow\qquad\vev{D_1(p)D_2(-p)}=\frac{4g_1g_2v^2}{p^2+m_V^2},}
where $m_V^2\equiv4(g_1^2+g_2^2)v^2$ is the squared mass of the vector multiplet. 
The diagram is then trivially evaluated in terms of the correlator function coefficient $\wt C_0(p^2)$ \Meade, and supersymmetry dictates the other contributions\foot{The assumption here is that the gauge multiplet masses are independent of SUSY breaking.  When the gaugino and gauge boson masses are split, we do not get a simple linear combination of the correlator coefficients \refs{\ggmgm,\Buican}.}.  Restoring group theory factors and defining $1/g^2\equiv1/g_1^2+1/g_2^2$, the result is
\eqn\mass{m_r^{2}=-g^4c_2(r)m_V^4\int\frac{d^4p}{(2\pi)^4}\frac1{p^2(p^2+m_V^2)^2}\left(\wt C_0(p^2)-4\wt C_{1/2}(p^2)+3\wt C_1(p^2)\right).}
For a scalar transforming in the representation $r$ of the group, $c_2(r)$ is the quadratic Casimir in that group and for that representation.  The ``diagonal'' nature of the breaking is such that the Casimir is the same before and after breaking.  For a product gauge group, one simply needs to sum over the factors with each having the above form.  Note that a different $m_V$ and and a different set of $\wt C$'s may appear for each group.  The above formula is a special case of one obtained in \McGb, which is the deconstructed \refs{\deca,\decb} version of \McGa. 

The hidden sector piece is familiar from ordinary gauge mediation, but the rest of the integrand is not.  One can relate this expression to more familiar results, however.  For example, one can use the following identity
\eqn\fid{\frac{m_V^4}{p^2(p^2+m_V^2)^2}=f(0)+f(m_V^2)-\frac2{m_V^2}\int_0^{m_V^2} dx~f(x),\qquad f(x)=\frac{p^2}{(p^2+x)^2}.}
This is useful because $f(m_V^2)$ is the factor in the integrand for the standard Higgsed case \comments, and $f(0)$ is that in the massless mediator case \Meade.  So for minimal gauge mediation \super, one can in principle use the known results for standard gauge mediation \refs{\dgp,\martin} and Higgsed gauge mediation \hgm.  The decomposition in \fid\ is not coincidental.  If one computes in the mass eigenbasis for the vector multiplets, there are three contributing diagrams corresponding to the three terms above; the first has two massless mediating fields, the second has two massive ones, and the third is from a mixed diagram.

The fact that the masses computed above are smaller than those that would result from a conventional mediation is readily demonstrated by considering the difference,
\eq{\frac1{p^2}-\frac{m_V^4}{p^2(p^2+m_V^2)^2}=\frac{p^2+2m_V^2}{(p^2+m_V^2)^2}>0.}
For a given hidden sector, we see that the integrand (and therefore the mass) from mediation via a massless vector field \Meade\ is always greater than that which we have computed.  

\bigskip
\bigskip
\noindent {\bf Acknowledgments:}

I benefitted from conversations with J. Evans, Z. Komargodski, Y. Ookouchi, and T.T. Yanagida.  
This work was supported by World Premier International Research Center Initiative (WPI
Initiative), MEXT, Japan. 

\listrefs
\bye